\documentclass[11pt,graphicx]{article}
\usepackage{amssymb,amsmath,amsfonts}
\usepackage{graphicx}
\usepackage{graphics}
\usepackage{eepic,epsfig}
\begin{document}
\title{Induced self-interactions in the spacetime of a global monopole with finite core}
\author{ E. R. Bezerra de Mello\thanks{E-mail: emello@fisica.ufpb.br}\\
Departamento de F\'{\i}sica-CCEN, Universidade Federal da Para\'{\i}ba\\
58.059-970, C. Postal 5.008, J. Pessoa, PB,  Brazil}
\maketitle

\begin{abstract}
In this paper we analyze induced self-interactions for point-like particles with electric and scalar charges placed at rest in the spacetime of a global monopole admitting a general spherically symmetric inner structure to it. In order to develop this analysis we calculate the three-dimensional Green function associated with the physical system under consideration. As we shall see for the charged particle outside the monopole core, the corresponding Green functions are composed by two distinct contributions, the firsts ones are induced by the non-trivial topology of the global monopole considered as a point-like defect and the seconds are corrections induced by the non-vanishing inner structure attributed to it. For both cases, the self-energies present a similar structure, having also two distinct contributions as well. For a specific model considered for region inside the monopole, named flower-pot, we shall see that the particle with electric charge will be always subject to a repulsive self-force with respect to the monopole core's boundary, on the other scalar charged particle exhibits peculiar behavior. Depending on the curvature coupling the self-force can be repulsive or attractive with respect to the core's boundary. Moreover, the contribution due to the point-like global monopole vanishes for massless particle conformally coupled with three dimensional space section of the manifold, and the only contribution comes from the core-induced part. 
\end{abstract}
\section{Introduction}
\label{Int}
It is well known that different types of topological objects may have been formed by the vacuum phase transition in the early Universe after Planck time \cite{Kibble,V-S}. These include domain walls, cosmic strings and
monopoles. Global monopoles are heavy spherically symmetric topological objects which may have been formed by the vacuum phase transition in the early Universe after Planck time. Although the global monopole was first introduced by Sokolov and Starobinsky in \cite{Soko}, its gravitational effects has been analyzed by Barriola and Vilenkin \cite{BV}. In the latter it is shown that for points far away from the monopole's center, the geometry of the spacetime can be given by the line element below:
\begin{equation}
ds^2=-dt^2+dr^2+\alpha^2r^2(d\theta^2+\sin^2\theta d\varphi^2)\ . 
\label{gm}
\end{equation}
In \eqref{gm} the parameter $\alpha^2$, which is smaller than unity, depends on the energy scale $\eta$ where the phase transition spontaneously occur. This spacetime has a non-vanishing scalar curvature, $R=\frac{2(1-\alpha^2)}{\alpha^2r^2}$ and presents a solid angle deficit $\delta\Omega= 4\pi^2(1-\alpha^2)$. 

Although the geometric properties of the spacetime outside the monopole are very well understood, there are no explicit expressions for the components of the metric tensor in the region inside. Consequently many interesting investigations of physical effects associated with global monopole consider this object as a point-like defect. Adopting this simplified model, many calculations of vacuum polarization effects associated with bosonic \cite{Lousto} and fermionic quantum fields \cite{Mello2}, in four-dimensional global monopole spacetime, present divergence on the monopole's core.

A very well known phenomenon that occur with an electric charged test particle placed at rest in a curved spacetime, is that it may become subjected to an electrostatic self-interactions. The origin of this induced self-interaction resides on the non-local structure of the field caused by the spacetime curvature and/or non-trivial topology. This phenomenon has been analyzed in an idealized cosmic string spacetime by Linet \cite{Linet} and Smith \cite{Smith}, independently, and also in the spacetime of a global monopole considered as a point-like defect in \cite{Mello3}. In these analysis, the corresponding self-forces are always repulsive; moreover they present divergences on the respective defects' core. A possible way to avoid the divergence problem is to consider these defects as having a non-vanishing radius, and attributing for the region inside a structure. For the cosmic string, two different models have been adopted to describe the geometry inside it: the ballpoint-pen model proposed independently by Gott and Hiscock \cite{Gott}, replaces the conical singularity at the string axis by a constant curvature spacetime in the interior region, and flower-pot model \cite{BA}, presents the curvature concentrated on a ring with the spacetime inside the string been flat.  Khusnutdinov and Bezerra in \cite{NV}, revisited the induced electrostatic self-energy problem considering the Hiscock and Gott model for the region inside the string. As to the global monopole the electrostatic self-energies problem have been analyzed considering for the region inside, the flower-pot model in \cite{Mello4} and ballpoint pen in \cite{Barbosa1}. In both analysis it was observed that the corresponding self-forces are finite at the monopole's core center. 

In the context of self-interactions the induced self-energy on scalar charged point-like particles on a curved spacetime reveals peculiarities \cite{Burko,Wiseman} due to the non-minimal curvature coupling with the geometry. In the case of of Schwarzschild spacetime, the self-force on a scalar charged particle at rest vanishes for minimal coupling \cite{Zelnikov}. The self-energy on scalar particle on the global monopole spacetime considering a non-trivial inner structure been developed recently in \cite{Barbosa}. 

In this present paper, mainly supported by two previous publications, \cite{Mello4,Barbosa}, we shall analyze the self-interaction problems associate with electric and scalar charged particles placed at rest in the global monopole spacetime, considering a non-trivial structure for the region inside to it. This paper is organized as follows: In section \ref{Model} we present the model to consider the geometry of the global in the whole space and the relevant field equations associated with the electric and scalar charged particles placed at rest in this background. We calculate the effective three-dimensional Green functions for points outside and inside the monopole's core. As a consequence, we provide a general expression for the electrostatic and scalar self-energies and their related self-forces. In section \ref{F-P} we calculate explicit expressions for the self-energies considering, as application of previous formalism, the flower-pot model for the region inside the monopole. Finally in section \ref{Conc}, we present our conclusions and more relevant remarks.  

\section{The system}\label{Model}
Many investigations concerning physical effects around a global monopole are developed considering it as idealized point-like defect. In this way the geometry of the spacetime is described by line element (\ref{gm}) for all values of the radial coordinate. However, a realistic model for a global monopole should present a non-vanishing characteristic core radius. For example, considering the model proposed by Barriola and Vilenkin \cite{BV}, the line element given by (\ref{gm}) is attained for the radial coordinate much lager than its characteristic core radius, which depends on the inverse of the energy scale where the global $O(3)$ symmetry is spontaneously broken to $U(1)$. Explicit expressions for the components of the metric tensor in whole space have not yet been found. Here, in this paper we shall not go into the details about this calculation. Instead, we shall consider a spherically symmetric model for describing the metric tensor for the region inside the shell of radius $a$. In the exterior region corresponding to $r>a$, the line element is given by (\ref{gm}), while in the interior region, $r<a$, the geometry is described by the static spherically symmetric line element
\begin{equation}
ds^{2}=-dt^{2}+v^{2}(r)dr^{2}+w^{2}(r)(d\theta ^{2}+\sin ^{2}\theta d\varphi ^{2}) \ .  
\label{gm1}
\end{equation}
Because the metric tensor must be continuous at the boundary of the core, the functions $v(r)$ and $w(r)$ must satisfy the conditions
\begin{equation}
v(a)=1 \ {\rm and} \ w(a)=\alpha a \ .  \label{bound}
\end{equation}

\subsection{Self-interactions}
In this subsection we shall develop a general formalism to analysis self-interactions associated with electric and scalar charged particles in global monopole spacetime. 

\subsubsection{Electrostatic self-interaction} 
Here we investigate the electrostatic self-energy and the self-force for a point-like charged particle at rest, induced by the spacetime geometry associated with a global monopole with the core of finite radius. We shall assume that in the region inside the monopole core the geometry is described by line element (\ref{gm1}), and in the exterior region we have the standard line
element (\ref{gm}). From the Maxwell equations, the covariant components of the electromagnetic four-vector potential, $A_\mu $, obey the equation below,
\begin{equation}
\partial _\lambda\left[ \sqrt{-g}g^{\mu\nu}g^{\lambda\sigma}\left( \partial _{\nu}A_{\sigma}-\partial_{\sigma}A_{\nu}\right) \right] =-4\pi \sqrt{-g}j^{\mu},  \label{Meq1}
\end{equation}
where $j^\mu$ is the four-vector electric current density. For a point-like particle at rest with coordinates $\vec{r}_{0}=(r_{0},\theta _{0},\varphi_{0})$, in the coordinate system corresponding to the line element (\ref{gm1}), the static four-vector current and potential are expressed by: $j^{\mu}=(j^{0},0,0,0)$ and $A_{\sigma}=(A_{0},0,0,0)$. The only nontrivial equation of (\ref{Meq1}) is $\mu=0$ with
\begin{equation}
j^{0}(x)=q\frac{\delta^{(3)}(\vec{r}-\vec{r}_{0})}{\sqrt{-g}}\ ,  \label{J0}
\end{equation}%
where $q$ is the electric charge of the particle. In the spherically symmetric spacetime defined by (\ref{gm1}), the differential equation obeyed by $A_{0}$ is:
\begin{equation}
\left[\partial _{r}\left( \frac{w^{2}}{uv}\partial _{r}\right) -\frac{v}{u}{\vec{L}}^{2}\right]A_{0}=-\frac{4\pi q}{\sin \theta }\delta (\vec{r}-\vec{r}_{0}) \ ,  \label{Meq2}
\end{equation}
with ${\vec{L}}$ being the operator orbital angular momentum. The solution of this equation can be written in terms of the Green function associated with the differential operator defined by the left-hand side, as
follows:
\begin{equation}
A_{0}(\vec{r})=4\pi qG(\vec{r},\vec{r}_{0})\ ,  \label{A0}
\end{equation}%
with the equation for the Green function
\begin{equation}
\left[ \partial _{r}\left( \frac{w^{2}}{uv}\partial _{r}\right) -\frac{v}{u}{\vec{L}}^{2}\right] G(\vec{r},\vec{r}_{0})=-\frac{\delta(r-r_{0})}{\sin \theta }\delta (\theta -\theta _{0})\delta (\varphi -\varphi
_{0}) \ .  \label{Green-a1}
\end{equation}

Having the electrostatic self-potential for the charge we can evaluate the corresponding self-force by using the standard formula
\begin{equation}
f_{el}^{i}(\vec{r}_{0})=-qg^{ik}F_{km}u^{m}=-q\frac{g^{ik}}{u}\partial _{k}A_{0}|_{\vec{r}=\vec{r}_{0}}=-4\pi q^{2}\frac{g^{ik}}{u}\lim_{\vec{r}\rightarrow \vec{r}_{0}}\left[ \partial _{k}G(\vec{r},%
\vec{r}_{0})\right] \ .  \label{feli}
\end{equation}
An alternative way to obtain the self-force is to consider first the electrostatic self-energy given by \cite{Linet,Smith}
\begin{equation}
U_{el}(\vec{r}_{0})=qA_{0}(\vec{r}_{0})/2=2\pi q^{2}\lim_{\vec{r}\rightarrow \vec{r}_{0}}G(\vec{r},\vec{r}_{0})\ ,
\label{SE0}
\end{equation}
and then to derive the force on the base of the formula%
\begin{equation}
f_{{el}}^{i}(\vec{r}_{0})=-\frac{g^{ik}}{u}\partial _{k}U_{{el}}(\vec{r}_{0}) \ .  \label{feli2}
\end{equation}

In (\ref{feli}) and (\ref{SE0}) the limit provides a divergent result. To obtain a finite and well defined result for the self-force, we should apply some renormalization procedure for the Green function. The procedure that we
shall adopt is the standard one: we subtract from the Green function the terms in the corresponding DeWitt-Schwinger adiabatic expansion which are divergent in the coincidence limit. So, we define the renormalized Green function as
\begin{equation}
G_{{ren}}(\vec{r},\vec{r}_{0})=G(\vec{r},\vec{r}_{0})-G_{Sing}(\vec{r},\vec{r}_{0})\ .  \label{Grenn}
\end{equation}
In this way the renormalized self-energy, $U_{{el,ren}}(\vec{r}_{0})$, and self-force, $f_{{el,ren}}^{i}(\vec{r}_{0})$, are obtained by the formulas (\ref{feli}) and (\ref{SE0}) substituting $G( \vec{r},\vec{r}_{0})$ by $G_{{ren}}(\vec{r},\vec{r}_{0})$. Note that here the subtraction of the divergent part of the Green function corresponds to the renormalization of the particle mass.

Taking into account the spherical symmetry of the problem, we may express the Green function by the ansatz below,
\begin{equation}
G(\vec{r},\vec{r}_{0})=\sum_{l=0}^{\infty}\sum_{m=-l}^{l}g_{l}(r,r_{0})Y_{l}^{m}(\theta ,\varphi )Y_{l}^{m\ast}(\theta _{0},\varphi _{0})\ ,  \label{Green-b1}
\end{equation}
with $Y_{l}^{m}(\theta ,\varphi )$ being the ordinary spherical harmonics. Substituting (\ref{Green-b1}) into (\ref{Green-a1}) and using the well known addition theorem for the spherical harmonics, we arrive at the differential equation for the unknown radial function:
\begin{equation}
\left[ \frac{d}{dr}\left( \frac{w^{2}}{uv}\frac{d}{dr}\right) -\frac{v}{u}l(l+1)\right] g_{l}(r,r_{0})=-\delta (r-r_{0}) \ .  \label{g}
\end{equation}%
As the functions $u(r)$, $v(r)$  and $w(r)$ are continuous at $r=a$, from (\ref{g}) it follows that the function $g_{l}(r,r_{0})$ and its first radial derivative are also continuous at this point. The function $g_{l}(r,r_{0})$ is also continuous at $r=r_{0}$. The discontinuity condition of the first radial derivative at $r=r_{0}$ is obtained by the integration of (\ref{g}) about this point. It reads,
\begin{equation}
\frac{dg_{l}(r,r_{0})}{dr}|_{r=r_{0}+}-\frac{dg_{l}(r,r_{0})}{dr}|_{r=r_{0}-}=-\frac{u(r_{0})v(r_{0})}{w^{2}(r_{0})} \ .  \label{derjunc}
\end{equation}

Let us denote by $R_{1l}(r)$ and $R_{2l}(r)$ the two linearly independent solutions of the homogeneous equation corresponding to (\ref{g}) in the region inside the monopole's core. We shall assume that the function $R_{1l}(r)$ is regular at the core center $r=r_{c}$ and that the solutions are normalized by the Wronskian relation
\begin{equation}
R_{1l}(r)R_{2l}^{\prime }(r)-R_{1l}^{\prime }(r)R_{2l}(r)=-\frac{u(r)v(r)}{w^{2}(r)} \ .  \label{Wronin}
\end{equation}%
In the region outside the core the linearly independent solutions to the
corresponding homogeneous equation are the functions $r^{\lambda _{1}}$ and $%
r^{\lambda _{2}}$, where
\begin{equation}
\lambda _{1,2}=-\frac{1}{2}\pm \frac{1}{2\alpha }\sqrt{\alpha ^{2}+4l(l+1)}\
\ .  \label{lam12}
\end{equation}%
Now, we can write $g_{l}(r,r_{0})$ as a function of the radial coordinate $r$ in the separate regions $[r_{c},\min (r_{0},a))$, $(\min (r_{0},a),\max(r_{0},a))$, and $(\max (r_{0},a),\infty )$ as a linear combination of the
above mentioned solutions with arbitrary coefficients. The requirement of the regularity at the core center and at the infinity reduces the number of these coefficients to four. They are determined by the continuity condition
at the monopole's core boundary and by the matching conditions at $r=r_{0}$. In this way we find the following expressions%
\begin{eqnarray}
g_{l}(r,r_{0}) &=&\frac{(ar_{0})^{\lambda _{1}}R_{1l}(r)}{\alpha ^{2}\left[aR_{1l}^{\prime }(a)-\lambda _{2}R_{1l}(a)\right] } \ ,\;\mathrm{for}\;r\leqslant a \ ,  \label{gl1in} \\
g_{l}(r,r_{0}) &=&\frac{r_{<}^{\lambda _{1}}r_{>}^{\lambda _{2}}}{\alpha^{2}(\lambda _{1}-\lambda _{2})}\left[ 1-\left( \frac{a}{r_{<}}\right)^{\lambda _{1}-\lambda _{2}}D_{1l}(a)\right] \  ,\;\mathrm{for}\;r\geqslant a \ ,
\label{gl1out}
\end{eqnarray}
in the case $r_{0}>a$, and
\begin{eqnarray}
g_{l}(r,r_{0})&=&R_{1l}(r_{<})R_{2l}(r_{>})-R_{1l}(r_{0})R_{1l}(r)D_{2l}(a) \ ,\;\mathrm{for} \;r\leqslant a \ ,  \label{gl2in} \\
g_{l}(r,r_{0}) &=&\frac{a^{\lambda _{1}}r^{\lambda _{2}}R_{1l}(r_{0})}{\alpha ^{2}\left[ aR_{1l}^{\prime }(a)-\lambda _{2}R_{1l}(a)\right] } \ ,\;\mathrm{for}\;r\geqslant a \ ,  \label{gl2out}
\end{eqnarray}%
in the case $r_{0}<a$. In these formulas, $r_{<}=\min (r,r_{0})$ and $r_{>}=\max (r,r_{0})$, and we have used the notation
\begin{equation}
D_{jl}(a)=\frac{aR_{jl}^{\prime }(a)-\lambda _{j}R_{jl}(a)}{aR_{1l}^{\prime}(a)-\lambda _{2}R_{1l}(a)} \ ,\;j=1,2 \ .  \label{Djl}
\end{equation}

First let us consider the case when the charge is outside the monopole's core ($r_{0}>a$). Substituting the function (\ref{gl1out}) into (\ref{Green-b1}), we observe that the Green function is presented in the form of the sum of two terms
\begin{equation}
G(\vec{r},\vec{r}_{0})=G_{{m}}(\vec{r},\vec{r}_{0})+G_{{c}}(\vec{r},\vec{r}_{0}) \ ,  \label{Ggumar}
\end{equation}%
where
\begin{equation}
G_{{m}}(\vec{r},\vec{r}_{0})=\frac{1}{4\pi \alpha r_{>}}\sum_{l=0}^{\infty }\frac{2l+1}{\sqrt{\alpha ^{2}+4l(l+1)}}\frac{r_{<}^{\lambda _{1}}}{r_{>}^{\lambda _{1}}}P_{l}(\cos \gamma ) \ ,  \label{Gm}
\end{equation}%
is the Green function associated with the geometry of a point-like global monopole, and the term
\begin{equation}
G_{{c}}(\vec{r},\vec{r}_{0})=-\frac{1}{4\pi \alpha }\sum_{l=0}^{\infty }\frac{(2l+1)D_{1l}(a)}{\sqrt{\alpha ^{2}+4l(l+1)}}(rr_{0})^{\lambda _{2}}P_{l}(\cos \gamma )\ \ ,  \label{Gc1}
\end{equation}%
is induced by non-trivial structure of the core. In formulas (\ref{Gm}) and (\ref{Gc1}), the $\gamma $ is the angle between the directions $(\theta,\varphi )$ and $(\theta _{0},\varphi _{0})$, and $P_{l}(x)$ represents the Legendre polynomials. As we can see, the contribution (\ref{Gc1}) depends on the structure of the core through the radial function $R_{1l}(r)$.

As we have already mentioned, the induced self-energy is obtained from the renormalized Green function taking the coincidence limit. We can observe that for points with $r>a$, the core-induced term (\ref{Gc1}) is finite in the coincidence limit and the divergence appears in the point-like monopole part only. So, in order to provide a finite and well defined value to (\ref{SE0}), we have to renormalize Green function $G_{{m}}(\vec{r},\vec{r
}_{0})$ only:
\begin{equation}
G_{{ren}}(r_{0},r_{0})=G_{{m,ren}}(r_{0},r_{0})+G_{{c}%
}(r_{0},r_{0})\ .  \label{Gren}
\end{equation}%
As explained before, to obtain $G_{{m,ren}}(r_{0},r_{0})$, we subtract from (\ref{Gm}) the terms in the corresponding DeWitt-Schwinger adiabatic expansion which are divergent in the coincidence limit:%
\begin{equation}
G_{{m,ren}}(r_{0},r_{0})=\lim_{\vec{r}\rightarrow \vec{r}_{0}}\left[ G_{{m}}(\vec{r},\vec{r}_{0})-G_{Sing}(\vec{r},\vec{r}_{0})\right] \ .  \label{Gmren}
\end{equation}%
The part $G_{Sing}(\vec{r},\vec{r}_{0})$ is found from the general formula given, for instance, in \cite{Chris}. For simplicity, taking the separation of the points along the radial direction only ($\gamma =0$), we find
\begin{equation}
G_{Sing}(r,r_{0})=\frac{1}{4\pi |r-r_{0}|}\ .
\label{Gmdiv}
\end{equation}
Now, by using formulas (\ref{Gm}) and (\ref{Gmdiv}), one obtains
\begin{equation}
G_{{m,ren}}(r_{0},r_{0})=\frac{1}{4\pi r_{0}}\lim_{t\rightarrow 1}\left[ \frac{1}{\alpha }\sum_{l=0}^{\infty }\frac{2l+1}{\sqrt{\alpha^{2}+4l(l+1)}}\ t^{\lambda _{1}}-\frac{1}{1-t}\right] ,  \label{Gmren1E}
\end{equation}%
where $t=r_{<}/r_{>}$. To evaluate the limit on the right, we note that
\begin{equation}
\lim_{t\rightarrow 1}\left( \frac{1}{\alpha }\sum_{l=0}^{\infty }t^{l/\alpha+1/2\alpha -1/2}-\frac{1}{1-t}\right) =0 \ .  \label{RelLim}
\end{equation}%
On the basis of this relation, replacing $1/(1-t)$ in (\ref{Gmren1E}) by the first term in the brackets in (\ref{RelLim}), we find
\begin{equation}
G_{{m,ren}}(r_{0},r_{0})=\frac{S(\alpha )}{4\pi r_{0}} \ ,
\label{Gmren2-1}
\end{equation}%
where we have introduced the notation%
\begin{equation}
S(\alpha )=\frac{1}{\alpha }\sum_{l=0}^{\infty }\left[ \frac{2l+1}{\sqrt{\alpha ^{2}+4l(l+1)}}-1\right] \ .  \label{falfa}
\end{equation}
The function $S(\alpha )$ is positive (negative) for $\alpha <1$ ($\alpha >1$) and, hence, the corresponding self-force is repulsive (attractive).

Combining formulas (\ref{SE0}), (\ref{Gc1}) and (\ref{Gmren2-1}), for the renormalized electrostatic self-energy we obtain
\begin{equation}
U_{{el,ren}}(r_{0})=\frac{q^{2}S(\alpha )}{2r_{0}}-\frac{q^{2}}{2\alpha r_{0}}\sum_{l=0}^{\infty }\frac{(2l+1)D_{1l}(a)}{\sqrt{\alpha^{2}+4l(l+1)}}\left( \frac{a}{r_{0}}\right) ^{\sqrt{1+4l(l+1)/\alpha ^{2}}} \
.  \label{U-out}
\end{equation}
As we can see, the second term of the renormalized self-energy provides a convergent series for $r_{0}>a$. Moreover, the dependence of the self-energy on the core structure is present in the function $D_{1l}(a)$. For large distances from the core, $r_{0}\gg a$, the main contribution into the core-induced part comes from
the term $l=0$ and one has%
\begin{equation}
U_{{el,ren}}(r_{0})\approx \frac{q^{2}}{2r_{0}}\left[ S(\alpha )-\frac{aD_{10}(a)}{\alpha ^{2}r_{0}}\right] \ .  \label{Ulargedist}
\end{equation}%
The self-force is obtained from (\ref{U-out}) by using formula (\ref{feli2}):%
\begin{equation}
\vec{f}_{{el,ren}}(\vec{r}_{0})=U_{{el,ren}}(r_{0})\frac{\vec{r}_{0}}{r_{0}^{2}}-\frac{q^{2}\vec{r}_{0}}{2\alpha ^{2}r_{0}^{3}}\sum_{l=0}^{\infty }(2l+1)D_{1l}(a)\left( \frac{a}{r_{0}}\right) ^{\sqrt{1+4l(l+1)/\alpha ^{2}}} \ .  \label{felout}
\end{equation}%
According to the symmetry of the problem, the self-force has only a radial component. 

Now let us study the case when the charge is inside the core, $r_{0}<a$. The corresponding Green function is obtained from (\ref{gl2in}) and is written in the form
\begin{equation}
G(\vec{r},\vec{r}_{0})=G_{0}(\vec{r},\vec{r}_{0})+G_{\alpha }(\vec{r},\vec{r}_{0}) \ ,  \label{Gin}
\end{equation}%
where
\begin{equation}
G_{0}(\vec{r},\vec{r}_{0})=\frac{1}{4\pi }\sum_{l=0}^{\infty}(2l+1)R_{1l}(r_{<})R_{2l}(r_{>})P_{l}(\cos \gamma ) \ ,  \label{G0in}
\end{equation}%
is the Green function for the background geometry described by the line element (\ref{gm1}) for all values $r_{c}\leqslant r<\infty $, and the term
\begin{equation}
G_{\alpha }(\vec{r},\vec{r}_{0})=-\frac{1}{4\pi }\sum_{l=0}^{\infty}(2l+1)R_{1l}(r_{0})R_{1l}(r)D_{2l}(a)P_{l}(\cos \gamma ) \ ,  \label{Galfain}
\end{equation}
is due to the global monopole geometry in the region $r>a$. For the points away from the core boundary the latter is finite in the coincidence limit. The self-energy for the charge inside the core is written in the form
\begin{equation}
U_{el,ren}(r_{0})=2\pi q^{2}G_{0,ren}(r_{0},r_{0})-\frac{q^{2}}{2}\sum_{l=0}^{\infty }(2l+1)D_{2l}(a)R_{1l}^{2}(r_{0}) \ ,
\label{Urenin}
\end{equation}
where
\begin{equation}
G_{0,ren}(r_{0},r_{0})=\lim_{\vec{r}\rightarrow \vec{r}_{0}}\left[ G_{0}(\vec{r},\vec{r}_{0})-G_{Sing}(\vec{r},\vec{r}_{0})\right] .  \label{G0renin}
\end{equation}%
The only contribution in the divergent part of the Green function comes from the first term of the DeWitt-Schwinger expansion. Note that near the center of the core one has $R_{1l}(r_{0})\propto (r_{0}-r_{c})^{l}$ and the main
contribution into the second term on the right of (\ref{Urenin}) comes from the term with $l=0$. Substituting the self-energy given by (\ref{Urenin}) into formula (\ref{feli2}), we obtain the self-force for the charge inside the monopole core.

\subsubsection{Scalar self-interaction}
The action associated with a charged massive scalar field, $\phi$, coupled with a charge density, $\rho$, in a curved background spacetime reads 
\begin{eqnarray}
\label{Action}
S=-\frac12\int \ d^4x \ \sqrt{-g} \ \left(g^{\mu\nu}\nabla_\mu\phi\nabla_\nu\phi+\xi R\phi^2+m^2\phi^2\right) +\int \ d^4x \ \sqrt{-g} \ \rho \ \phi \ ,
\end{eqnarray}
where the first part contains the Klein-Gordon action admitting an arbitrary curvature coupling, $\xi$, and the second part contains the interaction term. In the above equation $R$ represents the scalar curvature. The field equation can be obtained by varying the action with respect to the field. This provides
\begin{eqnarray}
\label{EM}
\left(\Box-\xi R-m^2\right)\phi=-\rho \ .
\end{eqnarray}

The physical system that we shall analyze corresponds to a particle at rest; so, there is no time dependence on the field. Moreover, because the metric tensor under consideration is also time-independent, the equation of motion above reduces effectively to a three-dimensional one:
\begin{eqnarray}
\label{EM1}
\left(\nabla^2-\xi R-m^2\right)\phi=-\rho \ .
\end{eqnarray}

The energy-momentum tensor associated with this system is obtained by taking the variation of (\ref{Action}) with respect to the metric tensor. It reads
\begin{eqnarray}
T_{\mu\nu}&=&\rho \ \phi \ g_{\mu\nu}+\nabla_\mu\phi\nabla_\nu\phi-\frac12g_{\mu\nu}\left(g^{\lambda\chi}\nabla_\lambda\phi\nabla_\chi\phi+m^2\phi^2 \right)\nonumber\\
&+&\xi\left(G_{\mu\nu}\phi^2+g_{\mu\nu}\Box\phi^2-\nabla_\mu\nabla_\mu\phi^2\right) \ ,
\end{eqnarray}
$G_{\mu\nu}$ being the Einstein tensor.

The energy for the scalar particle is obtained as shown below,
\begin{eqnarray}
E=-\int \ d^3x \ \sqrt{-g} \ T^0_0 \ .
\end{eqnarray}

For static fields configurations and by using the motion equation (\ref{EM1}), we have:
\begin{eqnarray}
\label{SE}
E=-\frac12\int \ d^3x \ \sqrt{-g} \  \rho \ \phi \ .
\end{eqnarray}
By using the three-dimensional Green function defined by the  differential operator in (\ref{EM1}),
\begin{eqnarray}
\label{Green}
\left(\nabla^2-\xi R-m^2\right)G({\vec{x}},{\vec{x}}')=-\frac{\delta^3({\vec{x}}-{\vec{x}}')}{\sqrt{-g}} \ ,
\end{eqnarray}
the energy given by \eqref{SE} can be written as
\begin{eqnarray}
\label{E1}
E=-\frac12\int\int \ d^3x \sqrt{-g(x)} \ d^3x' \sqrt{-g(x')} \  \rho({\vec{x}}) G({\vec{x}},{\vec{x}}') \ \rho({\vec{x}'}) \ .
\end{eqnarray}

Considering now a point-like scalar charge at rest at the point $\vec{r}_0$, the charge density takes the form,
\begin{eqnarray}
\label{Charge}
\rho(\vec{r})=\frac{q}{\sqrt{-g}}\delta^3({\vec{r}}-{\vec{r}}_0) \ .
\end{eqnarray}

Finally substituting (\ref{Charge}) into (\ref{E1}), we obtain for the energy the following expression:
\begin{eqnarray}
\label{E2}
E=-\frac{q^2}2G({\vec{r}_0},{\vec{r}_0})  \ .
\end{eqnarray}

Here also, the evaluation of the Green function that we need for the calculation of the energy is divergent at the coincidence limit. In order to obtain a finite and well defined result for the energy we subtract from the Green function the corresponding DeWitt-Schwinger asymptotic expansion. Following the general procedure given in \cite{Chris}, the singular behavior of the three-dimensional Green function associated with a massive scalar field reads:
\begin{equation}
G_{Sing}(x,x')=\frac1{4\pi}\left[\frac1{\sqrt{2\sigma}}-m\right]+O(\sigma) \ . \label{Had3D}
\end{equation}
Adopting the above mentioned renormalization approach, the scalar self-energy is given by
\begin{eqnarray}
\label{ERen}
E_{Ren}&=&-\frac{q^2}2G_{Ren}({\vec{r}_0},{\vec{r}_0})  \  ,\nonumber\\
&=&-\frac{q^2}2\lim_{{\vec{r}}\to{\vec{r}}_0}[G({\vec{r}},{\vec{r}_0})-G_{Sing}({\vec{r}},{\vec{r}_0})] \ .
\end{eqnarray}

Once more, taking into account the spherical symmetry of the problem, the scalar Green function can be expressed by the ansatz below,
\begin{equation}
G(\vec{r},\vec{r}_0)=\sum_{l=0}^\infty\sum_{m=-l}^lg_l(r,r_0)Y_l^m(\theta ,\varphi )Y_l^{m\ast}(\theta_0,\varphi_0)\ .  \label{Green-a}
\end{equation}
Substituting (\ref{Green-a}) into (\ref{Green}), we arrive at the following differential equation for the unknown radial function $g_l(r,r')$:
\begin{eqnarray}
\label{gr}
\left[\frac d{dr}\left(\frac{w^2}{v}\frac{d}{dr}\right)-l(l+1)v-\xi Rvw^2-m^2vw^2\right]g_l=-\delta(r-r_0) \ ,
\end{eqnarray}
with the Ricci scalar being given by
\begin{equation}
R=\frac2{w^2}+\frac{4v'w'}{wv^3}-\frac{4w''}{wv^2}-\frac{2(w')^2}{w^2v^2} \ . 
\end{equation}
For $\xi=0$ and $m=0$ the differential equation above is similar to the previous one found in the analysis of electrostatic self-energy, Eq. \eqref{g}. Moreover; as to the solution of (\ref{gr}) the junctions conditions are obtained as explain below: 
\begin{itemize}
\item Because the function $v(r)$ and $w(r)$ are continuous at $r=a$, it follows that $g_l(r,r')$ should be continuous at this point; however, due to the second radial derivative of the function $w(r)$ in the Ricci scalar, a Dirac-delta function contribution on the Ricci scalar takes place if the first derivative of this function is not continuous at the boundary.\footnote{For the flower-pot model this fact occur and has been considered in \cite{Mello4}.} Naming by $\check{R}=\bar{R}\delta(r-a)$ the Dirac-delta contribution of the Ricci scalar, the junction condition on the boundary is:
\begin{equation}
\frac{dg_l(r)}{dr}|_{r=a^+}-\frac{dg_l(r)}{dr}|_{r=a^-}=\xi\bar{R} \ g_l(a) \ . \label{Cond1}
\end{equation}
\item The function $g_l(r)$ is continuous at $r=r_0$, however by integrating (\ref{gr}) about this point, the first radial derivative of this function obeys the junction condition below: 
\begin{equation}
\frac{dg_l(r)}{dr}|_{r=r_0^+}-\frac{dg_l(r)}{dr}|_{r=r_0^-}=-\frac{v(r_0)}{w^2(r_0)} \ . \label{Cond2}
\end{equation}
\end{itemize}

Now after this general discussion, let us analyze the solutions of the homogeneous differential equation associated with (\ref{gr}) for regions inside and outside the monopole's core. Let us denote by $R_{1l}(r)$ and $R_{2l}(r)$ the two linearly independent solutions of the equation in the region inside with $R_{1l}(r)$ been the regular one at the core center $r=r_{c}$; moreover, we shall assume that solutions are normalized by the Wronskian relation \eqref{Wronin}.

In the region outside, the two linearly independent solution are:
\begin{equation}
\frac1{\sqrt{r}}I_{\nu_l}(mr) \ \ {\rm and} \ \frac1{\sqrt{r}}K_{\nu_l}(mr) \ ,
\end{equation}
where $I_\nu$ and $K_\nu$ are the modified Bessel functions of order
\begin{equation}
\nu_l=\frac1{2\alpha}\sqrt{(2l+1)^2+(1-\alpha^2)(8\xi-1)} \ .
\end{equation} 

Now we can write the function $g_l(r)$ as a linear combination of the above solutions with arbitrary coefficients for the regions $(r_{c}, \min (r_0,a))$, $(\min (r_0,a), \ \max (r_0,a))$, and $(\max (r_0,a), \infty)$. The requirement of the regularity at the core center and at the infinity reduces the number of these coefficients to four. These constants are determined by the continuity condition at the monopole's core boundary and at the point $r=r_0$ by the junctions conditions given in (\ref{Cond1}) and (\ref{Cond2}), respectively. In this way we find the following expressions:
\begin{eqnarray}
g_l(r,r_0)&=&\frac{K_{\nu_l}(mr_0)R_{1l}(r)}{\alpha^2\sqrt{ar_0}}\frac1{a{\cal{R}}_l^{(1)}(a)K_{\nu_l}(ma)-R_{1l}(a){\tilde{K}}_{\nu_l}(ma)}  \ \ {\rm for} \ r\leq a \ , {\label{gout-}}\\ 
g_l(r,r_0)&=&\frac{I_{\nu_l}(mr_<)K_{\nu_l}(mr_>)}{\alpha^2\sqrt{rr_0}}-D_l^{(+)}(a)\frac{K_{\nu_l}(mr)K_{\nu_l}(mr_0)}{\alpha^2\sqrt{rr_0}}\ \ {\rm for} \ r\geq a \ , \label{gout+}
\end{eqnarray}
in the case of the charged particle is outside the monopole, $r_0>a$ , and 
\begin{eqnarray}
g_l(r,r_0)&=&R_{1l}(r_<)[R_{2l}(r_>)-D_l^{(-)}(a)R_{1l}(r_>)] \ \ {\rm for} \ r\leq a \ , \label{gin-} \\ 
g_l(r,r_0)&=&\frac{R_{1l}(r_0)K_{\nu_l}(mr)}{\alpha^2\sqrt{ar}}\frac1{a{\cal{R}}_l^{(1)}(a)K_{\nu_l}(ma)-R_{1l}(a){\tilde{K}}_{\nu_l}(ma)} \ \ {\rm for} \ \ r\geq a \ , \label{gin+}
\end{eqnarray}
in the case of the charged particle is inside the monopole, $r_0<a$. In these formulas, $r_{<}=\min (r,r')$ and $r_{>}=\max (r,r')$, and we have used the notations:
\begin{eqnarray}
D_{l}^{(+)}(a)&=&\frac{a{\cal{R}}_{l}^{(1)}(a)I_{\nu_l}(ma)-R_{1l}(a){\tilde{I}}_{\nu_l}(ma)}{a{\cal{R}}_{l}^{(1)}(a)K_{\nu_l}(ma)-R_{1l}(a){\tilde{K}}_{\nu_l}(ma)}  \ , \label{D+} \\
D_l^{(-)}(a)&=&\frac{a{\cal{R}}_{l}^{(2)}(a)K_{\nu_l}(ma)-R_{2l}(a){\tilde{K}}_{\nu_l}(ma)}{a{\cal{R}}_{l}^{(1)}(a)K_{\nu_l}(ma)-R_{1l}(a){\tilde{K}}_{\nu_l}(ma)} \ . \label{D-}
\end{eqnarray}
For a given function  $F(z)$, we use the notation
\begin{eqnarray}
{\tilde{F}}(z)=zF'(z)-\frac12F(z) 
\end{eqnarray}
and for a solution $R_{jl}(r)$, with $j= 1, \ 2$,
\begin{equation}
\label{Rsing}
{\cal{R}}_l^{(j)}(a)=R_{jl}'(a)+\xi{\bar{R}}R_{jl}(a) \ .
\end{equation}
In the above definition $R'_{jl}(a)=\frac{dR_{jl}(r)}{dr}|_{r=a}$.

Before to go for a specific model, let us still continue the investigation of the self-energy for this general spherically symmetric spacetime. First we shall consider the case where the charge is outside the monopole's core. Substituting (\ref{gout+}) into (\ref{Green-a}) we see that the Green function is expressed in terms of two contributions:
\begin{equation}
G(\vec{r},\vec{r}_0)=G_{gm}(\vec{r},\vec{r}_0)+G_c(\vec{r},\vec{r}_0) \ , 
\end{equation}
where
\begin{eqnarray}
G_{gm}(\vec{r},\vec{r}_0)=\frac1{4\pi\alpha^2{\sqrt{rr_0}}}\sum_{l=0}^\infty (2l+1)I_{\nu_l}(mr_<)K_{\nu_l}(mr_>)P_l(\cos\gamma) \label{G-gm}
\end{eqnarray}
and 
\begin{eqnarray}
G_c(\vec{r},\vec{r}_0)=-\frac1{4\pi\alpha^2{\sqrt{rr_0}}}\sum_{l=0}^\infty (2l+1)D_l^{(+)}(a)K_{\nu_l}(mr_0) K_{\nu_l}(mr)P_l(\cos\gamma) \ . \label{Gc}
\end{eqnarray}
The first part corresponds to the Green function for the geometry of a point-like global monopole and the second is induced by the non-trivial structure of its core. In the formulas above, $\gamma$ is the angle between both directions $(\theta, \ \varphi)$ and  $(\theta_0, \ \varphi_0)$ and $P_l(x)$ represents the Legendre polynomials of degree $l$.

The induced scalar self-energy is obtained by taking the coincidence limit in the renormalized Green function. We can observe that for points with $r>a$, the core-induced term (\ref{Gc}) is finite and the divergence appears in the point-like monopole part only. So, in order to provide a finite and well defined result for (\ref{ERen}), we have to renormalize the Green function $G_{gm}(\vec{r},\vec{r}_0)$ only. Let us first take $\gamma=0$ in the above expressions. The renormalized Green function is expressed by:
\begin{equation}
G_{Ren}(r_0,r_0)=G_{gm,ren}(r_0,r_0)+G_c(r_0,r_0) \ ,  \label{Gren}
\end{equation}
where 
\begin{equation}
G_{gm,ren}(r_0,r_0)=\lim_{r\to r_0}[G_{gm}(r,r_0)-G_{Sing}(r,r_0)] \ .
\end{equation}
For points outside the core, the radial one-half of geodesic distance becomes $|r-r_0|/2$, we have
\begin{equation}
G_{Sing}(r,r_0)=\frac1{4\pi}\left[\frac1{|r-r_0|}-m\right] \ .
\end{equation}

Now, by using $G_m(r_0,r)$ given in (\ref{G-gm}), we have:
\begin{equation}
G_{gm,ren}(r_0,r)=\frac1{4\pi r_0}\lim_{r\to r_0}\left[\frac1{\alpha^2}\sum_{l=0}^\infty{(2l+1)}I_{\nu_l}(mr_<)K_{\nu_l}(mr_>)-\frac{1}{1-t}\right]+\frac{m}{4\pi} ,  \label{Gmren1}
\end{equation}
where $t=r_{<}/r_{>}$. In order to evaluate the limit on the right hand side of the above equation, we take the identity \eqref{RelLim}. So, as consequence of this relation and replacing in (\ref{Gmren1}) the expression $1/(1-t)$ by the first term in the brackets of that identity, we find
\begin{equation}
G_{{gm,ren}}(r',r')=\frac{S_{(\alpha )}(mr')}{4\pi\alpha r}+\frac{m}{4\pi} \ ,
\label{Gmren2}
\end{equation}
with
\begin{equation}
S_{(\alpha)}(mr')=\sum_{l=0}^\infty\left[\frac{(2l+1)}{\alpha}I_{\nu_l}(mr')K_{\nu_l}(mr')-1\right] \label{S} \ .
\end{equation}

Two special situations deserve to be analyzed:
\begin{itemize}
\item For the case where $\xi=0$, we have $\nu_l=\frac1{2\alpha}{\sqrt{4l(l+1)+\alpha^2}}$. Taking the limit $m\to 0$ in (\ref{S}), using \cite{Grad,Abra} we get a position independent expression named $S_\alpha$:
\begin{equation}
S_{\alpha}=\sum_{l=0}^\infty\left[\frac{2l+1}{\sqrt{4l(l+1)+\alpha^2}}-1\right] \ .
\end{equation}
Up to a factor $1/\alpha$ this expression coincides with the similar one obtained in the previous analysis of the electrostatic self-energy, Eq \eqref{falfa}.
\item For $\xi=1/8$ we have $\nu_l=\frac{1}{2\alpha}({2l+1})$. Taking the limit $m\to 0$ in (\ref{S}) we see that the term inside the bracket vanishes, consequently $G_{{gm,ren}}(r',r')=0$ and the only contribution to the scalar self-energy comes from the core-induced part (\ref{Gc}). In fact under these circumstance, the differential equation obeyed by the Green function in the region outside the monopole is conformally related with the corresponding one in a flat space due to the conformal flatness of the space section of this metric tensor:
\begin{equation}
d{\vec{l}}^2=dr^2+\alpha^2r^2d\Omega_{(2)}=\rho^\lambda(d\rho^2+\rho^2d\Omega_{(2)}) \ ,
\end{equation}
with $\rho=(\alpha r)^{1/\alpha}$ being $\lambda=2(\alpha-1)$. Moreover, by explicit calculation we can show that $G_{gm}(\vec{r},\vec{r}')=\rho^{-\lambda/4}G_M (\vec{\rho},\vec{\rho}')\rho'^{-\lambda/4}$.
\end{itemize}

Finally the complete expression for scalar self-energy reads
\begin{eqnarray}
E_{Ren}=-\frac{q^2}{8\pi\alpha r_p}S_{(\alpha)}(mr_p)-\frac{q^2m}{8\pi}+\frac{q^2}{8\pi\alpha^2r_p} \sum_{l=0}^\infty(2l+1)D_l^{(+)}(a)\left(K_{\nu_l}(mr_p)\right)^2 \ . \label{TE}
\end{eqnarray}

The self-force on a static test particle can be calculated by taking the negative gradient of the corresponding self-energy \cite{Burko},
\begin{equation}
\vec{f}={\vec{\nabla}}E_{Ren} \ .
\end{equation} 
Considering $\vec{f}=f_r\hat{r}$, the radial component of this force reads:
\begin{eqnarray}
f_r&=&\frac{q^2}{8\pi\alpha r_p^2}S_{(\alpha)}(mr_p)-\frac{q^2m}{8\pi\alpha^2r_p}\sum_{l=0}^\infty(2l+1)\left[\left(I_{\nu_l+1}(mr_p)+ \frac{\nu_l I_{\nu_l+1}(mr_p)}{mr_p}\right) K_{\nu_l}(mr_p)\right.\nonumber\\
&-&\left.I_{\nu_l}(mr_p)\left(K_{\nu_l+1}(mr_p)-\frac{\nu_lK_{\nu_l}(mr_p)}{mr_p}\right)\right]-\frac{q^2}{8\pi\alpha^2r^2_p} \sum_{l=0}^\infty(2l+1)D_l^{(+)}(a)\left(K_{\nu_l}(mr_p)\right)^2\nonumber\\
&-&\frac{q^2m}{4\pi\alpha^2r_p}\sum_{l=0}^\infty(2l+1)D_l^{(+)}(a)K_{\nu_l}(mr_p)\left(K_{\nu_l+1}(mr_p)-\frac{\nu_lK_{\nu_l}(mr_p)}{mr_p}\right) \ .
\end{eqnarray}

The second analysis that can be formally developed here, is related with the case when the charge is inside the core. The corresponding Green function can be written in the form
\begin{equation}
G(\vec{r},\vec{r}_0)=G_{0}(\vec{r},\vec{r}_0)+G_\alpha(\vec{r},\vec{r}_0) \ , 
\end{equation}
where
\begin{equation}
G_{0}(\vec{r},\vec{r}_0)=\frac{1}{4\pi }\sum_{l=0}^{\infty}(2l+1)R_{1l}(r_<)R_{2l}(r_>)P_{l}(\cos \gamma ) \ ,  \label{G0inS1}
\end{equation}
is the Green function for the background geometry described by the line element (\ref{gm1}) and the term
\begin{equation}
G_{\alpha}(\vec{r},\vec{r}_0)=-\frac{1}{4\pi }\sum_{l=0}^{\infty}(2l+1)D_l^{(-)}(a)R_{1l}(r)R_{1l}(r_0)P_{l}(\cos \gamma ) \ ,  \label{GalfainS1}
\end{equation}
is due to the global monopole geometry in the region $r>a$. For the points away from the core boundary the latter is finite in the coincidence limit.
The renormalized scalar self-energy for the charge inside is written in the form
\begin{equation}
\label{ERin}
E_{Ren}=-\frac{q^2}2G_{0,Ren}(\vec{r}_0,\vec{r}_0)+\frac{q^{2}}{8\pi}\sum_{l=0}^{\infty }(2l+1)D_l^{(-)}(a)(R_{1l}(r_0))^2,
\end{equation}
where the renormalized Green function is given by
\begin{equation}
G_{0,Ren}(\vec{r}_0,\vec{r}_0)=\lim_{\vec{r}\to\vec{r}_p}\left[G_{0}(\vec{r},\vec{r}_0)-G_{Sing}(\vec{r},\vec{r}_0)\right] \  .  \label{G0renin}
\end{equation}
Because the divergent part of the Green function should have the same structure as (\ref{Had3D}), the above expression provides a finite result; moreover, notice that near the center of the core one has $R_{1l}(r)\approx\frac{I_{l+1/2}(m(r-r_{c}))}{\sqrt{r-r_c}}$ and the main
contribution into the second term on the right of (\ref{ERin}) comes from the term with $l=0$. Finally we can say that the self-force is again obtained by taking the negative gradient of (\ref{ERin}).

\section{Flower-pot model}
\label{F-P}
As we have mentioned before, there is no closed expression for the metric tensor in the region inside the global monopole. However, adopting the flower-pot model for this region, the calculations of vacuum polarization effects associated with massive scalar and fermionic fields have been developed in \cite{Mello5,Mello6}, respectively. So, motived by these result we decided, as an illustration of the general procedure described before, to consider the flower-pot model in the present analysis of the induced electrostatic and scalar self-interactions. For this model the interior line element has the form \cite{Mello4}
\begin{equation}
ds^{2}=-dt^{2}+dr^{2}+\left[ r+(\alpha -1)a\right] ^{2}(d^{2}\theta +\sin^{2}\theta d^{2}\varphi ) \ .  \label{intflow}
\end{equation}
In terms of the radial coordinate $r$ the origin is located at $r=r_{c}=(1-\alpha )a$. Defining $\tilde{r}=r+(\alpha -1)a$, the line element takes the standard Minkowskian form. As we have mentioned before, from the Israel matching conditions for the metric tensors corresponding to (\ref{gm}) and (\ref{intflow}), we find the singular contribution for the scalar curvature located on the bounding surface $r=a$ \cite{Mello5}:
\begin{equation}
\bar{R}=4\frac{(1-\alpha)}{\alpha a} \ .
\label{Rbar}
\end{equation}

In what follows, we shall consider, separately, the analysis of electrostatic and scalar self-interactions.

\subsection{Electrostatic self-interaction}
Now we can express the renormalized Green function in the region outside the monopole core by taking into account that in the interior region we have two linearly independent solutions of the homogeneous equation corresponding to \eqref{g}:
\begin{equation}
R_{1l}(r)=\tilde{r}^{l} \ , \ {\rm and} \ R_{2l}(r)=\tilde{r}^{-l-1}/(2l+1) \ .
\label{R12Flower}
\end{equation}%
So, from formula (\ref{U-out}), the self-energy in the exterior region reads
\begin{eqnarray}
U_{el,ren}(r_{0}) &=&\frac{q^{2}S(\alpha )}{2r_{0}}+\frac{2q^{2}(1-\alpha )}{\alpha r_{0}}\sum_{l=0}^{\infty }\frac{l(2l+1)}{\sqrt{\alpha ^{2}+4l(l+1)}}  \nonumber\\
&&\times \frac{(a/r_{0})^{\sqrt{1+4l(l+1)/\alpha ^{2}}}}{\left[ \sqrt{\alpha^{2}+4l(l+1)}+\alpha +2l\right] ^{2}} \ .  \label{UrenFlow}
\end{eqnarray}
The second term on the right of this formula is positive for $\alpha <1$ and negative for $\alpha >1$. Combining this with the properties of the function $S(\alpha )$ discussed in the previous section, we conclude that the
electrostatic self-energy is positive for $\alpha <1$ and negative for $\alpha >1$. The corresponding self-force is directly found from (\ref{felout}) and is repulsive in the first case and attractive in the second one. The core-induced part in (\ref{UrenFlow}) diverges at the core boundary, $r_{0}=a$. Noting that for points near the boundary the main contribution into (\ref{UrenFlow}) comes from large values $l$, to the leading order we find
\begin{equation}
U_{\mathrm{el,ren}}(r_{0})\approx q^{2}\frac{(\alpha -1)}{8\alpha a}\ln \left[1-\left( a/r_{0}\right) ^{1/\alpha }\right]  \ , \label{GrenFlow}
\end{equation}
and the self-energy is dominated by the core-induced part.

Now we turn to the investigation of the self-energy for the particle inside the monopole core. Substituting the functions (\ref{R12Flower}) into formulas (\ref{G0in}) and (\ref{Galfain}), for the
corresponding Green functions in the interior region one finds%
\begin{eqnarray}
G_{0}(\mathbf{r},\mathbf{r}_{0}) &=&\frac{1}{4\pi |\mathbf{r}-\mathbf{r}_{0}|} \ , \label{G0Flower} \\
G_{\alpha }(\mathbf{r},\mathbf{r}_{0}) &=&\frac{1}{4\pi \alpha a}\sum_{l=0}^{\infty }\frac{2l+2-\alpha -\sqrt{\alpha ^{2}+4l(l+1)}}{2l+\alpha+\sqrt{\alpha ^{2}+4l(l+1)}}\frac{(\tilde{r}{_{0}}\tilde{r})^{l}}{(\alpha
a)^{2l}}P_{l}(\cos \gamma ) \ ,  \label{GalfaFlower}
\end{eqnarray}
Because in the flower-pot model the geometry in the region inside the monopole is a Minkowski one, we have $G_{sing}(\vec{r},\vec{r}_{0})=G_{0}(\vec{r},\vec{r}_{0})$ and, hence, $G_{0,ren}(r_{0},r_{0})=0$. Finally, the electrostatic self-energy in the region inside the monopole core reads:
\begin{equation}
U_{\mathrm{el,ren}}(r_{0})=2\pi q^{2}G_{\mathrm{ren}}(r_{0},r_{0})=\frac{q^{2}}{2\alpha a}\sum_{l=0}^{\infty }\frac{2l+2-\alpha -\sqrt{\alpha^{2}+4l(l+1)}}{2l+\alpha +\sqrt{\alpha ^{2}+4l(l+1)}}\left( \frac{\tilde{r}{_{0}}}{\alpha a}\right) ^{2l} \ .  \label{Gr-in1}
\end{equation}%
As in the case of the exterior region, this self-energy is positive for $\alpha <1$ and negative for $\alpha >1$. The corresponding self-force is easily found from relation (\ref{feli2}) and is repulsive with respect to
the boundary of the monopole core in the first case and attractive in the second case. Near the core center the main contribution into the self-energy comes from the lowest modes and one has%
\begin{equation}
U_{el,ren}(r_{0})\approx \frac{q^{2}}{2\alpha a}\left[ \frac{1-\alpha }{\alpha }+\frac{4-\alpha -\sqrt{\alpha ^{2}+8}}{2+\alpha +\sqrt{\alpha ^{2}+8}}\left( \frac{\tilde{r}{_{0}}}{\alpha a}\right) ^{2}\right] \  .
\label{Unearcent}
\end{equation}

\subsection{Scalar self-interaction}

In the region inside the global monopole the two linearly independent solutions for the homogeneous radial equation corresponding to \eqref{gr} are:
\begin{eqnarray}
R_{1l}(r)=\frac{I_{l+1/2}(m\tilde{r})}{\sqrt{\tilde{r}}} \ {\rm and} \ R_{2l}(r)=\frac{K_{l+1/2}(m\tilde{r})}{\sqrt{\tilde{r}}}  \ . \label{Flower-Int}
\end{eqnarray}
Having obtained the above solutions, the expressions for the Green functions in both, inside and outside regions, can be explicitly constructed, consequently the corresponding self-energies. These expressions depend on the coefficients $D^{(+)}_l(a)$ and $D^{(-)}_l(a)$, which can be explicitly provided as shown below:
\begin{eqnarray}
\label{D+-}
D^{(+)}_l(a)=\frac{n^{(+)}_l(a)}{d_l(a)}  \ \ {\rm and} \ \ D^{(-)}_l(a)=\frac{n^{(-)}_l(a)}{d_l(a)} \ ,
\end{eqnarray}
with
\begin{eqnarray}
\label{n+}
n^{(+)}_l(a)&=&I_{\nu_l}(ma)I_{l+1/2}(m\alpha a)\left[\frac{(l+4\xi(1-\alpha))}{\alpha}-\nu_l+\frac12\right]\nonumber\\
&+&ma\left[I_{\nu_l}(ma)I_{l+3/2}(m\alpha a) -I_{l+1/2}(m\alpha a)I_{\nu_l+1}(ma)\right] \ ,
\end{eqnarray}
\begin{eqnarray}
\label{n-}
n^{(-)}_l(a)&=&K_{\nu_l}(ma)K_{l+1/2}(m\alpha a)\left[\frac{(l+4\xi(1-\alpha))}{\alpha}-\nu_l+\frac12\right]\nonumber\\
&-&ma\left[K_{\nu_l}(ma)K_{l+3/2}(m\alpha a) -K_{l+1/2}(m\alpha a)K_{\nu_l+1}(ma)\right] 
\end{eqnarray}
and
\begin{eqnarray}
\label{d}
d_l(a)&=&K_{\nu_l}(ma)I_{l+1/2}(m\alpha a)\left[\frac{(l+4\xi(1-\alpha))}{\alpha}-\nu_l+\frac12\right]\nonumber\\
&+&ma\left[K_{\nu_l}(ma)I_{l+3/2}(m\alpha a) +I_{l+1/2}(m\alpha a)K_{\nu_l+1}(ma)\right] \ .
\end{eqnarray}

As in the first analyze, let us consider the core-induced part of the self-energy for the region outside, Eq. ({\ref{TE}}), adopting specific values for the parameter $\xi$ and mass of the particle. Taking $\xi=0$, and by using \cite{Abra} to obtain the behavior of all functions contained in the coefficient defined in (\ref{D+-}), (\ref{n+}) and (\ref{d}), in the limit $m\to 0$, we have:
\begin{eqnarray}
D^{(+)}_l(a)\approx\frac{2}{\nu_l}\left(\frac{ma}2\right)^{2\nu_l}\frac{(\alpha+2l-2\alpha\nu_l)}{(2\alpha\nu_l+\alpha+2l)(\Gamma(\nu_l))^2} \ .
\end{eqnarray}
So the general term inside the summation of the core-induced part reads,
\begin{eqnarray}
\frac{\alpha(2l+1)}{\sqrt{4l^2+4l+\alpha^2}}\frac{4l(\alpha-1)}{(\sqrt{4l^2+4l+\alpha^2}+2l+\alpha)^2}\left(\frac ar\right)^{\sqrt{1+4l(l+1)/\alpha^2}} \ .
\end{eqnarray}
Which coincides with the result obtained for the electrostatic case.

We can see that the core-induced part in ({\ref{TE}}) is divergent near the boundary $r=a$. In order to verify this fact, we should analyze the general term in the summation for large value of $l$. Taking again the uniform asymptotic expansion for large orders of the modified Bessel functions in $D^{(+)}(a)$ and the same for the Macdonald function, $K_{\nu_l}(mr)$, we get,
\begin{eqnarray}
\label{Approx1}
-\frac{[\alpha-2\alpha\nu_l+2l+8\xi(1-\alpha)]}{[\alpha+2\alpha\nu_l+2l+8\xi(1-\alpha)]}\left(\frac ar\right)^{2\nu_l} \ .
\end{eqnarray}
At this point we want to mention that the contribution proportional to the curvature coupling, $\xi$, in the above expression is consequence of the delta-Dirac contribution in the Ricci scalar, given by $\xi R$, present in (\ref{Rsing}). 

For large value of $l$ we have $2\alpha\nu_l\approx(2l+1)+\frac{(1-\alpha^2)(8\xi-1)}{2(2l+1)}+O\left(\frac1{(2l+1)^2}\right)$. Substituting this expansion into (\ref{Approx1}), for the leading term in $1/l$, we obtain
\begin{eqnarray}
-\alpha(1-\alpha)\frac{(1-8\xi)}{4l}\left(\frac ar\right)^{2l/\alpha} \ .
\end{eqnarray}
Finally, after some intermediate steps we find:
\begin{equation}
\label{FP+}
E_{Ren}\approx q^2\frac{(1-\alpha)(1-8\xi)}{32\pi\alpha a}\ln\left[1-\left(a/r_0\right)^{1/\alpha}\right] \ .
\end{equation}
We can see that the above result does not depend on the mass of the particle. In fact this happens because the leading order term in the expansions of the coefficient $D^{(+)}(a)$ there appear a power factor $(ma)^{2\nu_l}$, as to the square of the Macdonald function $K_{\nu_l}(mr)$, the leading power factor in the mass is $(mr)^{-2\nu_l}$. Moreover we can see from the above result that for $\xi=1/8$ there is no divergent contribution in the core-induced part, and that for $\xi>1/8 $ this contribution becomes negative.

Now let us turn our investigation of the self-energy for the region inside  the monopole. Substituting the functions (\ref{Flower-Int}) into the formulas (\ref{G0inS1}) and (\ref{GalfainS1}) for the corresponding Green function in the interior region one finds \cite{Abra},
\begin{eqnarray}
\label{G00}
G_{0}(\vec{r},\vec{r}')=\frac1{4\pi}\frac{e^{-mR}}{R} \ ,
\end{eqnarray}
with $R=\sqrt{(\tilde{r}')^2+(\tilde{r})^2-2\tilde{r}\tilde{r}'\cos\gamma}$, being $\gamma$ the angle between the two directions defined by the unit vectors $\hat{\tilde{r}}'$ and $\hat{\tilde{r}}$. Taking $\gamma=0$ we get $R=|r-r'|$. Because in the flower-pot model the geometry in the region inside the monopole is a Minkowski one, we have $G_{0}(\vec{r},\vec{r}')=G_{Sing}(\vec{r},\vec{r}')$, consequently $G_{0,Ren}(\vec{r}_0,\vec{r}_0)=0$. The scalar self-energy in this region is given only by the core-induced part:
\begin{equation}
\label{E-ind}
E_{Ren}=\frac{q^2}{8\pi{\tilde{r}}_p}\sum_{l=0}^\infty(2l+1)D^{(-)}_l(a)\left(I_{l+1/2}(m\tilde{r}_0)\right)^2 \ ,
\end{equation}
being $\tilde{r}_0=r_0+(\alpha-1)a$. Near the core's center, $\tilde{r}_0\approx 0$, 
\begin{eqnarray}
I_{l+1/2}(m\tilde{r}_0)\approx\left(\frac{m\tilde{r}_0}{2}\right)^{l+1/2}\frac1{\Gamma(l+3/2)} \ , 
\end{eqnarray}
so the main contribution into the self-energy comes from the lowest mode, $l=0$, resulting in
\begin{equation}
\label{EE}
E_{Ren}\approx \frac{q^2mD_0^{(-)}(a)}{4\pi^2} \ .
\end{equation}
Taking the expression for the coefficient $D_l^{(-)}(a)$, for $l=0$, and considering $\xi=0$, after some steps we find
\begin{equation}
D_0^{(-)}(a)=\frac12\frac{\pi e^{-m\alpha a}(\alpha-1)}{\sinh(m\alpha a)(\alpha+\alpha am-1)+\alpha am\cosh(m\alpha a)} \ .
\end{equation}
Finally substituting the above expression into (\ref{EE}), and taking the limit $m\to 0$ we obtain
\begin{equation}
E_{Ren}(r)\approx \frac{q^2(\alpha-1)}{8\pi\alpha^2a} \ .
\end{equation}

Also we can calculate the core-induced contribution on the scalar self-energy near the boundary. Again, adopting a similar procedure as in the previous corresponding analysis, we can verify after some intermediate steps that the leading term inside the summation in (\ref{E-ind}) behaves as,
\begin{equation}
\frac1{4l}\frac{(1-\alpha)(8\xi-1)}{\alpha a}\left(\frac {\tilde{r}}{\alpha a}\right)^{2l} \ .
\end{equation}
Finally, taking this expression back into (\ref{E-ind}) we obtain,
\begin{equation}
\label{FP-}
E_{Ren}\approx q^2\frac{(1-\alpha)(1-8\xi)}{32\pi\alpha a}\ln\left(1-\frac{\tilde{r}_0}{\alpha a}\right) \ .
\end{equation} 
Once more we can see that this divergent contribution vanishes for $\xi=1/8$. 

Before to finish this application we want to cal attention that (\ref{FP+}) and (\ref{FP-}), for $\xi=0$, coincide, up the numerical factor $4\pi$, with the corresponding core-induced electrostatic self-energies derived in \cite{Mello4}.

\section{Concluding remarks}
\label{Conc}
In this paper we have analyzed induced self-energies associated with particle with electric and scalar charges place at rest in the global monopole spacetime, considering a inner structure to it. These two distinct situations have been investigated separately along this paper. 

As we could see, both investigations depend on the corresponding three-dimensional Green functions. For the general spherically symmetric static model of the core with finite thickness we have constructed the corresponding Green functions in both exterior and interior regions. In the region outside the core these functions are presented as a sum of two distinct contributions. The firsts ones corresponds to the Green functions for the geometry of a point-like global monopole and the seconds ones are induced by the non-trivial structure of the monopole core. A similar structure is also presented by the Green functions for the region inside the monopole. 

The self-energies are formally expressed in terms of the evaluation of the respective Green functions in the coincidence limit; however, because the results are divergent, we had to apply some renormalization procedure in order to obtain finites and well defined values. Here we used the point-splitting procedure. We analyze the divergent contributions associated with the Green functions at the coincidence limit, and extract all of them. In fact we did this in a manifest form by subtracting from the Green functions the DeWit-Schwinger adiabatic expansions which are divergent in the coincidence limit. 

As an application of the general results, in section \ref{F-P} we have considered a simple core model with a flat spacetime, the so called flower-pot model. 

For the electrostatic case, the corresponding self-forces are repulsive with respect to the core boundary in the case $\alpha <1$ and attractive for $\alpha >1$. In particular, for the first case, the charge placed at the core center is in a stable equilibrium position.  Although in the flower-pot model, we have found a finite value of the self-energy at the monopole's center, it presents a logarithmic singular behavior at the core boundary. 

As to the scalar charged particle the main conclusions found in this work, three deserves to be mentioned. They are: $i)$ The renormalized self-energy depends strongly on the value adopted for the curvature coupling constant $\xi$. For specific values of this constant, the self-energy may provide repulsive, or attractive self-forces with respect to the boundary. $ii)$ The self-energy presents a finite value at the monopole's core center, and $iii)$ for confomally coupled massless field, the self-energy only depends on the core-iduced part. 
\section*{Acknowledgment}
The author thanks CNPq. for partial financial.

\end{document}